\documentclass[twocolumn]{aastex631}
\usepackage{amssymb}
\usepackage{natbib}
\usepackage{hyperref}
\usepackage{breakurl}
\usepackage{xcolor}
\usepackage{enumitem}
\setlist{leftmargin=5.5mm}
\newcommand{\mail}{lilirayhk@phys.ncku.edu.tw}

\newcommand{\pflux}{\,ph\,cm$^{-2}$\,s$^{-1}$}

\newcommand{\kms}{\,km\,s$^{-1}$}

\newcommand\T{\rule{0pt}{2.6ex}}
\newcommand\B{\rule[-1.2ex]{0pt}{0pt}}
\newcommand{\nova}{ASASSN-16ma}
\shorttitle{Gamma-Ray Pulsations from \nova}
\shortauthors{Li et al.}

\begin{document}
\title{Evidence for Gamma-Ray Pulsations from the Classical Nova \nova}

\correspondingauthor{Kwan-Lok Li}
\email{\mail}

\author[0000-0001-8229-2024]{Kwan-Lok Li}
\affiliation{Department of Physics, National Cheng Kung University, 70101 Tainan, Taiwan}

\begin{abstract}
We report here a new result extracted from the \textit{Fermi} Large Area Telescope observation of the classical nova \nova\ that exhibits coherent $\gamma$-ray pulsations at 544.84(7) seconds during its outburst in 2016. Considering the number of independent trials, the significance of the evidence is 4.0$\sigma$, equivalent to a false alarm probability of $5.9\times10^{-5}$. The periodicity was steady during the 4 days of its appearance, indicating its origin as the spinning signal of the white dwarf. Given that the optical and $\gamma$-ray light curves of some shock-powered $\gamma$-ray novae have been recently shown closely correlated to each other, the $\gamma$-ray pulsation phenomenon likely implies an existence of the associated optical pulsations, which would provide detailed ephemerides for these extreme white dwarf binaries for further investigations in the near future.
\end{abstract}

\section{Introduction}
Discovered by the All Sky Automated Survey for SuperNovae (ASAS-SN; \citealt{2014ApJ...788...48S}) in October 2016 \citep{2016ATel.9669....1S}, \nova\ (also known as V5856 Sgr) was a 5.4-mag classical nova, which could be observed even with the naked eye in dark skies. At the same time, it was also a powerful $\gamma$-ray transient detected by the \textit{Fermi} Large Area Telescope (LAT) 14 days after the nova discovery \citep{2017NatAs...1..697L}. The highest $\gamma$-ray flux reached $F_{\rm 0.1-300GeV}\approx10^{-6}$\pflux\ on the first day of the LAT detection, and this peak level is among the brightest $\gamma$-ray novae \citep{2010Sci...329..817A,2014Sci...345..554A,2017NatAs...1..697L,2020NatAs...4..776A}. More important, the $\gamma$-ray and optical light curves of \nova\ closely tracked each other during the 9-day $\gamma$-ray active phase (from MJD~57700.0 to 57709.0). The correlation strongly suggests that both the emission components originated from shocks, showing that a nova can be shock-powered entirely for the first time \citep{2017NatAs...1..697L}.

Motivated by the recent discovery of the $\sim$500-s X-ray pulsations of the Nova Her 2021 (also known as V1674~Her; \citealt{2021ATel14720....1M,2021ATel14776....1M,2021ApJ...922L..42D}), we re-analysed the \textit{Fermi}-LAT data of \nova\ to study its fast $\gamma$-ray variability on a similar time-scale. We chose \nova\ as the primary target in this pilot search, because its $\gamma$-ray flux was one of the brightest and the daily $\gamma$-ray LAT light curve is relatively steady compared to others \citep{2017NatAs...1..697L}. These features provide a good chance for such a search. 

In this Letter, we present the search method used for \nova, a possible coherent signal found in the \textit{Fermi}-LAT observations, a physical interpretation that could explain the timing signal and its caveats, and the implication of the $\gamma$-ray pulsations in the future.

\begin{figure*}
\includegraphics[width=0.8\textwidth]{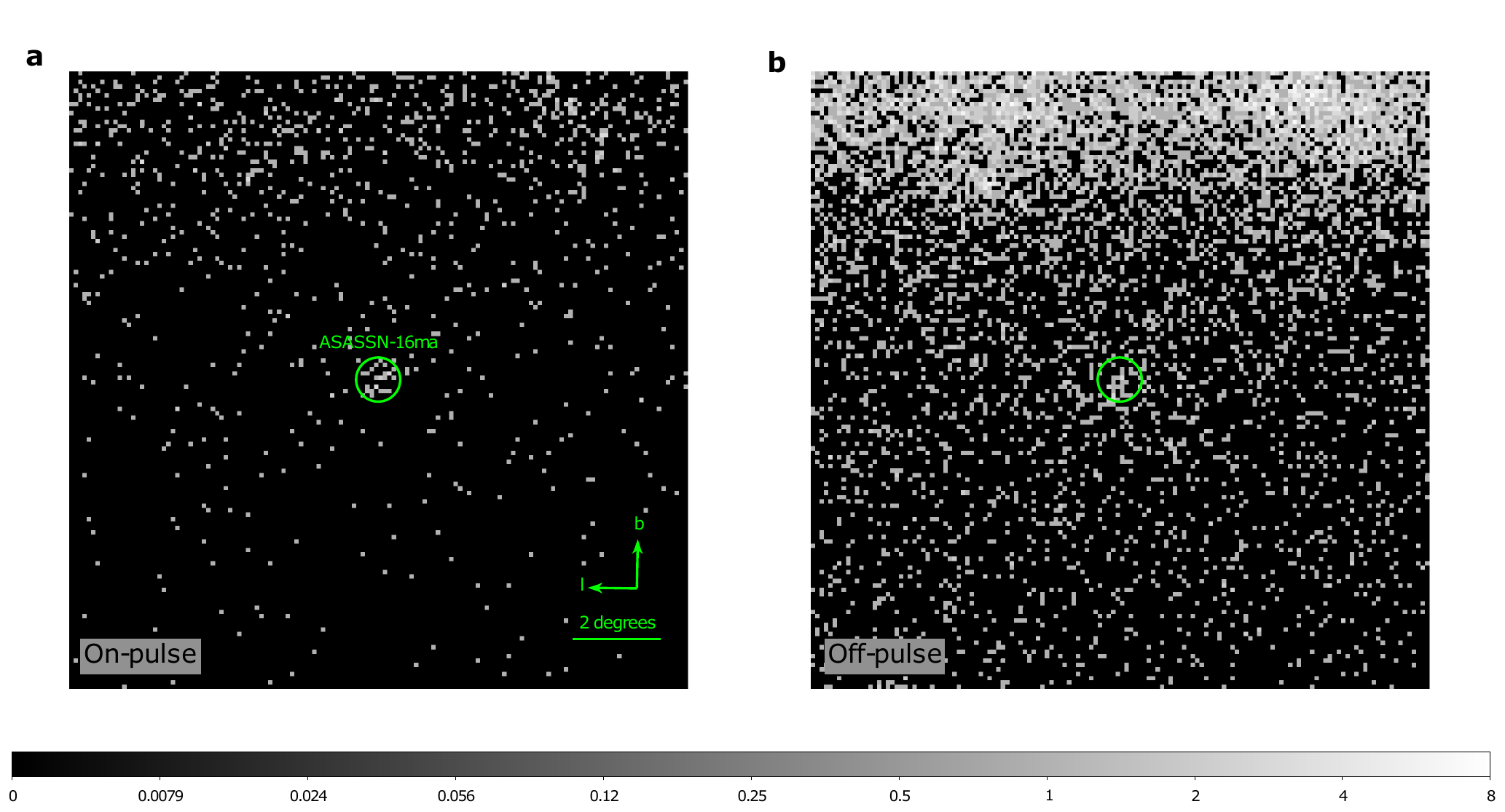}
\centering
\caption{\textit{Fermi}-LAT count maps of \nova\ during the on-pulse (a) and off-phase (b) intervals. The on/off-pulse internals were defined in the main text. The green circular region indicates the 0\fdg5 radius aperture for the source count extraction.
}
\label{fig:countmap}
\end{figure*}

\section{Period Searches on the \textit{Fermi}-LAT data}
We downloaded the relevant \textit{Fermi}-LAT observation from the \textit{Fermi}-LAT data Server at the \textit{Fermi} Science Support Center (FSSC) to investigate the timing behaviour of \nova\ during the 9 days of the $\gamma$-ray active period (MJD~57700.0--57709.0, during which the daily detection significances were all above $3\sigma$). The data version is Pass 8 Release 3 Version 3 (i.e., \texttt{P8R3\_SOURCE\_V3}), and the energy range is from 100~MeV to 300~GeV. The LAT-dedicated analysis software package, \texttt{Fermitools} (version 2.0.8), with the associated data files, \texttt{Fermitools-Data} (version 0.18), were used to perform all the \textit{Fermi}-LAT data analyses presented in the Letter, including the phase-resolved likelihood analysis described in a later section.

The task \texttt{gtselect} was used to extract the source event list, and hence, the aperture light curve of \nova. All the selected photons are in the event class of 128 (the recommended class for standard analyses) and the event type of 3 (collected in the front or back section of the LAT tracker).
In addition, the zenith angle must be lower than 90$\arcdeg$ to minimise the contamination from the Earth's limb.
As the nova was located near the Galactic plane, the background due to the Galactic diffusion emission is large. For a good balance between an effective source count extraction and a low background count contamination (the nova is just a few degrees off from the Galactic plane), an aperture radius of 0\fdg5 was adopted. Note that, although the LAT point spread function (PSF) is larger than 0\fdg5 below 1~GeV, the sensitivity of LAT is much higher in 1--100~GeV, in which the PSF shrinks to a sub-degree level. The choice of the aperture size can be well justified by the on-pulse count map produced later in the analysis process, which shows that the majority of the source counts were inside the aperture (Figure \ref{fig:countmap}a).
The event list was further filtered by \texttt{gtmktime} based on the Good Time Intervals (GTIs). The GTIs were defined by: (i) the data quality flag must be larger than zero to eliminate bad data, particle events, and events associated with solar flares (i.e., \texttt{DATA\_QUAL>0}); and (ii) the LAT configuration flag must equal to 1 to ensure the detector was operated in a regular science mode (i.e., \texttt{LAT\_CONFIG==1}). The final event list was barycentric corrected by \texttt{gtbary}.

\begin{figure*}
\includegraphics[width=0.8\textwidth]{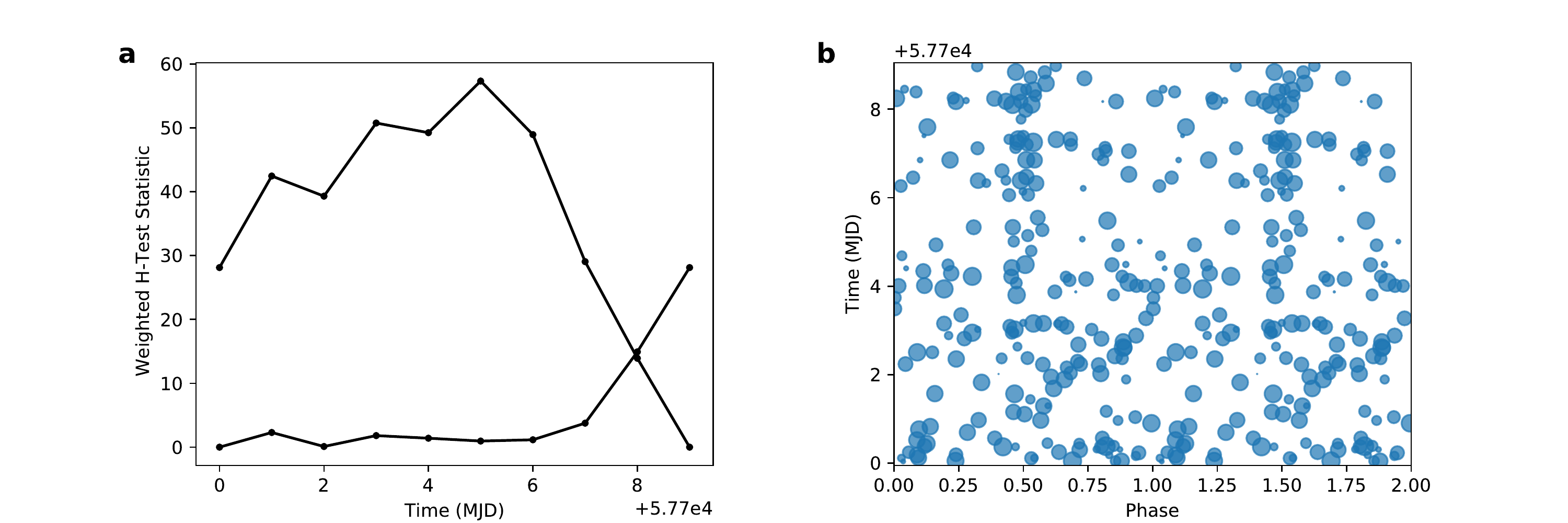}
\includegraphics[width=0.8\textwidth]{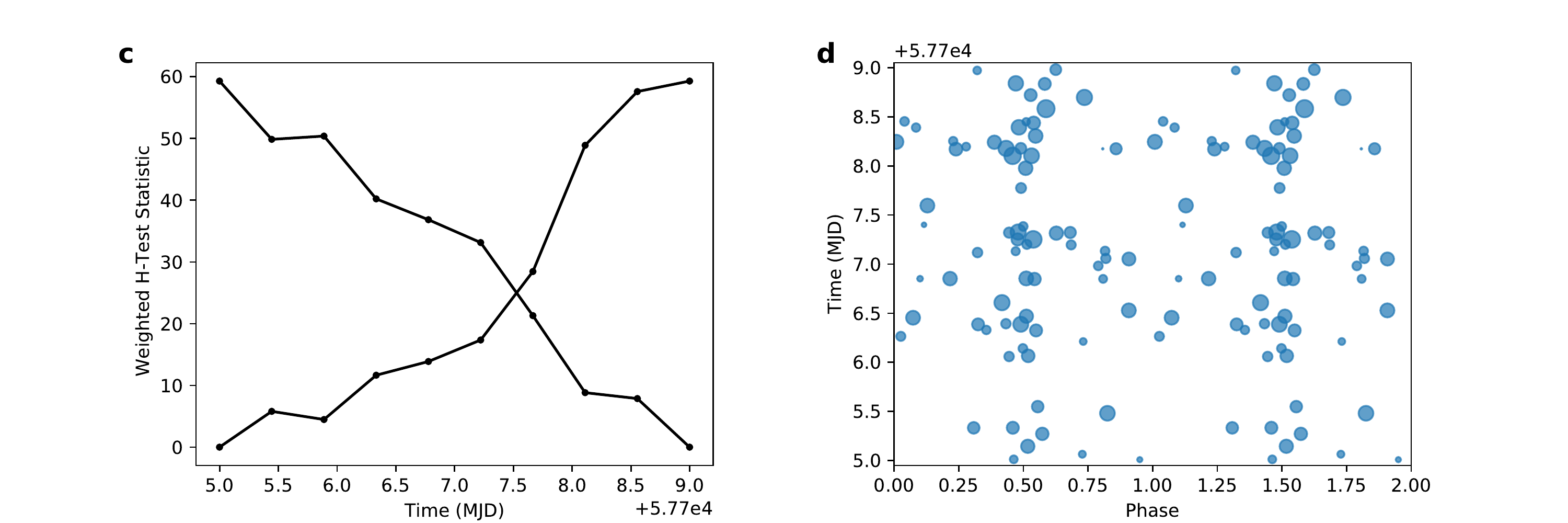}
\centering
\caption{Time evolution of the 544.84-s periodicity in $\bm{\gamma}$-rays. Top left (a): The weighted H-test statistic ($H_w$) of the 544.84-s periodicity accumulated from the beginning to the end of the observation (from MJD~57700.0 to 57709.0), and in reverse order. Both curves start from $H_w=0$.
Top right (b): The time distribution of the LAT-detected $\gamma$-ray photons phased according to the 544.84-s ephemeris. Each circle represents a photon count, and the size of the circle reflects the weight of the event applied in the $H_w$ calculation. The largest circle in the figure has a weight of about 1.
Bottom panel (c and d): The same set of plots as the top panel, but the time interval is cut down to MJD~57705.0--57709.0, during which the 544.84-s signal is persistent. The resultant test statistic is $H_w=59.3$, equivalent to a single-trial two-tailed significance of 5.5$\sigma$. In b and d, the $\gamma$-ray pulse is shifted to phase $=0.5$ for a good visualization, and two cycles are shown for clarity.}
\label{fig:htest}
\end{figure*}

\begin{figure*}
\includegraphics[width=0.8\textwidth]{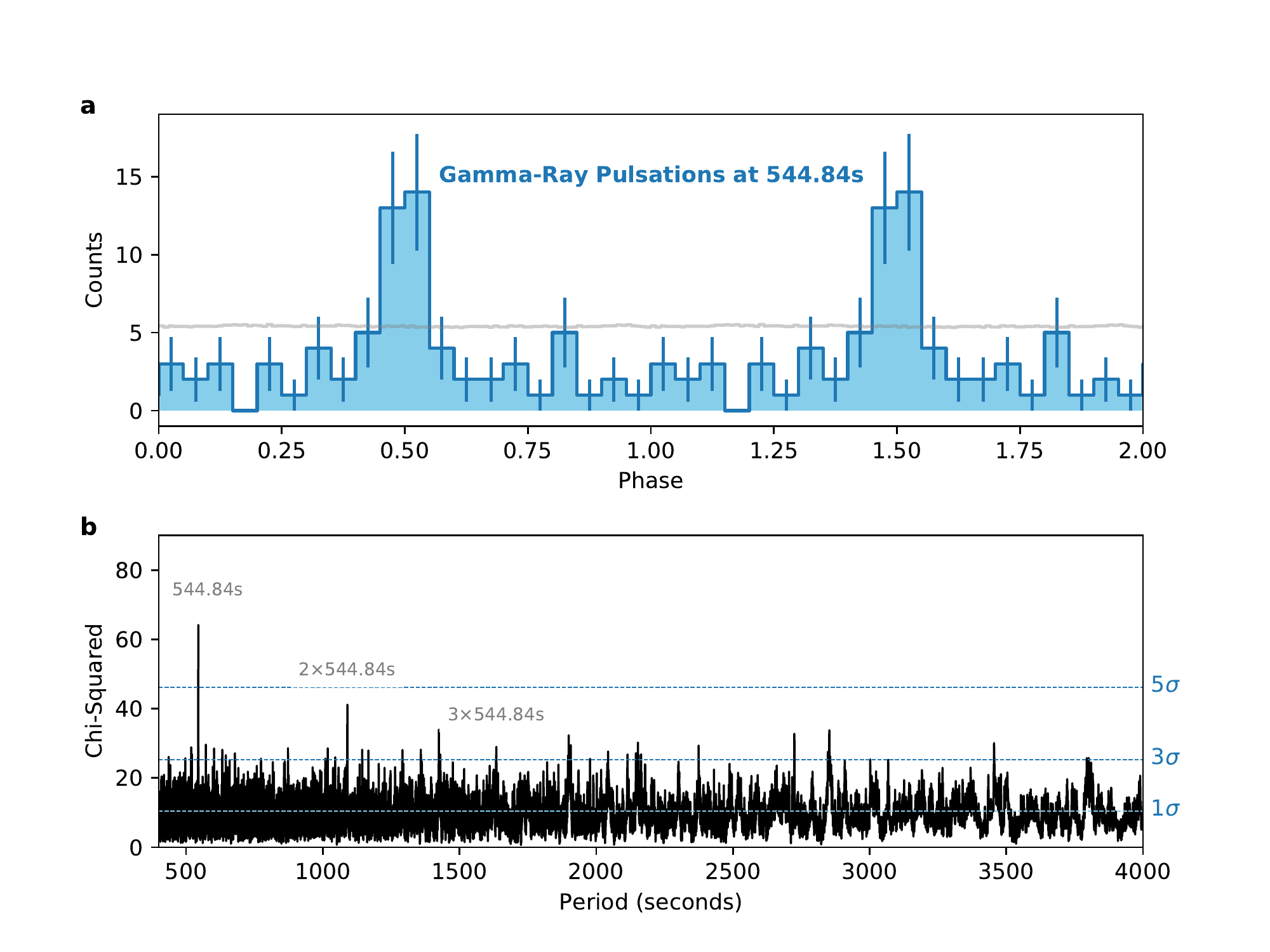}
\centering
\caption{The $\gamma$-ray pulse profile and the periodogram of \nova.
Top panel (a): The phased \textit{Fermi}-LAT aperture light curve of \nova\ using $P=544.84\,$s. The phase zero is defined at MJD~57704.99779 to make the $\gamma$-ray pulses locate around phase 0.5 for a better visualization. The grey line indicates the effective exposures of the LAT observations at different phases. The scale of the exposure curve is arbitrary, but it is clear that the detected $\gamma$-ray pulsations are not an effect of uneven sampling. All the reported errors are 1-$\sigma$ uncertainties, and two cycles are shown in the plot for clarity.
Bottom panel (b): The $\chi^2$-statistic periodogram of the same $\gamma$-ray dataset as (a). In a resolution of 0.1~s, the 544.84-s periodicity is significantly detected. In addition, the signals associated with the multiples of the period can also be seen. Using the $\chi^2$ distribution, the detection significance of the major signal is better than 5$\sigma$, in agreement with the H-test result (Figure \ref{fig:htest}).}
\label{fig:phased_lc}
\end{figure*}

The 9-day source event list was then processed with the \texttt{efsearch} task in the \texttt{HEASoft} package (version 6.28) to search for periodic signals using the $\chi^2$ test. As mentioned in the previous section, the search was motivated by the $\sim$500-s X-ray pulsation detected in Nova Her 2021 \citep{2021ATel14776....1M,2021ApJ...922L..42D}, and we therefore performed a test run from 400~s to 600~s in a resolution of 0.1~s, which surprisingly gives a marginal detection at 544.9~s. Nevertheless, other prominent but weaker spikes are also present in the $\chi^2$-statistic periodogram. To clarify the statistical significance, the weighted H-test statistic (\citealt{1989A&A...221..180D,2011ApJ...732...38K,2019A&A...622A.108B}; $H_w$) was employed. The associated test statistic is defined as

\begin{eqnarray}
\nonumber
H_w = \max_{1 \leqslant i \leqslant m}[Z_{iw}^2-4(i-1)],
\end{eqnarray}

where $Z_{iw}^2$ is the weighted statistic of the $Z_m^2$ test \citep{2011ApJ...732...38K} and $m=20$ is the maximum harmonic considered \citep{1989A&A...221..180D}. The event weights were computed for the LAT data in 0\fdg5 radius aperture using the user-contributed code, \texttt{add\_weights}, obtained from the FSSC, in which the LAT PSF is considered \citep{2019A&A...622A.108B}.
In the calculation, \texttt{TEMPO2} \citep{2006MNRAS.369..655H} with the \textit{Fermi} plug-in was used to convert the mission elapsed times (METs) into phases. This gives $H_w\approx30$, equivalent to a significance of $\approx2\sigma$ \citep{2019A&A...622A.108B}.

For a further analysis, we investigated the evolutions of $H_w$ forward and backward in time. While the two methods give the same value of $H_w\approx30$ at the end, the reverse evolutionary track yields a high peak of $\approx60$ on MJD~57705 (Figure \ref{fig:htest}a). The regular track starts rising around the same time also. These strongly suggest that the periodic signal began on MJD~57705. Therefore, we focused on this interval (i.e., MJD~57705.0--57709.0) fine-tuning the period in the range from $P=544.7\,$s to $544.9\,$s with a step of 0.01~s. An improved signal was found at $P=544.84\,$s (Figure \ref{fig:phased_lc}a), with which the regular and reverse tracks are both strictly growing over the whole interval (Figure \ref{fig:htest}c) with a maximum value of $H_w=59.3$. The corresponding detection significance can be inferred by \cite{2019A&A...622A.108B},
\begin{eqnarray}
\log P(H_w>x) = -3.80655 + \lambda_1(W)(x-22) \;\; {\rm for}\;x>29\nonumber
\end{eqnarray}
and
\begin{eqnarray}
\nonumber
\lambda_1(W)=-0.173025+0.0525796e^{-(W+5)/215.17}\\
+0.086406e^{-(W+5)/35.5709},\nonumber
\end{eqnarray}
where $W=30.24$ is the sum of the weights in the dataset. These result in $P(H_w>59.3)=4.0\times10^{-8}$, equivalent to a single-trial significance of 5.5$\sigma$ (two-tailed). For an independent check as well as a wider range of search, a $\chi^2$-statistic periodogram from $P=$~400~s to 4000~s of a resolution of 0.1~s was produced and the 544.84-s periodicity is the only signal detected beyond 5$\sigma$ (Figure \ref{fig:phased_lc}).

The trials factor of the whole analysis was estimated to be 1473, which includes 650 independent Fourier bins in the 400--600-s $\chi^2$-statistic search, 781 independent Fourier bins in the 400--4000-s $\chi^2$-statistic search, 20 attempts for the \texttt{add\_weights} code optimizing the parameter for the event weighting, 21 attempts for obtaining the best period in the time range from $P=544.7\,$s to $544.9\,$s with a step of 0.01~s, and 1 attempt for the time interval cut. The factor would increase the false alarm probability to $P_{\rm trials}=1-[1-P(H_w>59.3)]^{1473}=5.9\times10^{-5}$, equivalent to a significance of 4.0$\sigma$ (two-tailed). Given that some trails are correlated (e.g., between the first and the second $\chi^2$-statistic searches) and some attempts are actually optional (e.g., the second $\chi^2$-statistic search was extended to 4000~s for the completeness of the study, rather than a signal detection), the computed significance should be considered conservative. 

According to FSSC, the 544.84-s periodicity is not a known instrumental or orbital effect of the \textit{Fermi} satellite. The LAT background emission was examined using a random circular region of a radius of 2$\arcdeg$ (4.6 degrees away from the nova), and no periodicity can be found with the exact same approach for \nova. A 1-s-binned LAT exposure curve for \nova\ was computed by \texttt{gtexposure}. The phased curve stayed flat over the period, indicating the detected pulses are not due to a sampling effect (Figure \ref{fig:phased_lc}a). All the evidence suggests that the origin of the 544.84-s periodicity is astrophysical.

\subsection{Uncertainty Estimation of the 544.84-s Periodicity}
The uncertainty estimation for periods is usually non-trivial, and an ad hoc approach was applied to estimate the 1-$\sigma$ uncertainty of the 544.84-s periodicity.
According to Figure \ref{fig:phased_lc} (extracted using the 0.5-degree aperture), the $\gamma$-ray emission pulsed only in 10\% of the time, and 27 photons were detected in this interval. As the period is shifted from the best-fit, the profile becomes flatter and the peak decreases. Assuming that the expected peak of the profile is indeed 27, the 1-$\sigma$ measurement deviation allowed will be $\pm \sqrt{27}\approx\pm5$ (Poisson distribution was assumed). To measure the upper and lower boundaries of the period where the pulse drops to the 1-$\sigma$ level, a test period was shifted from the best-fit in a 0.01~s step. We found that the 1-$\sigma$ confidence interval of the periodicity is 544.79--544.91~s, and therefore set the larger upper deviation, $\Delta P=0.07\,$s, as the 1-$\sigma$ uncertainty.

\begin{figure}
\includegraphics[width=0.5\textwidth]{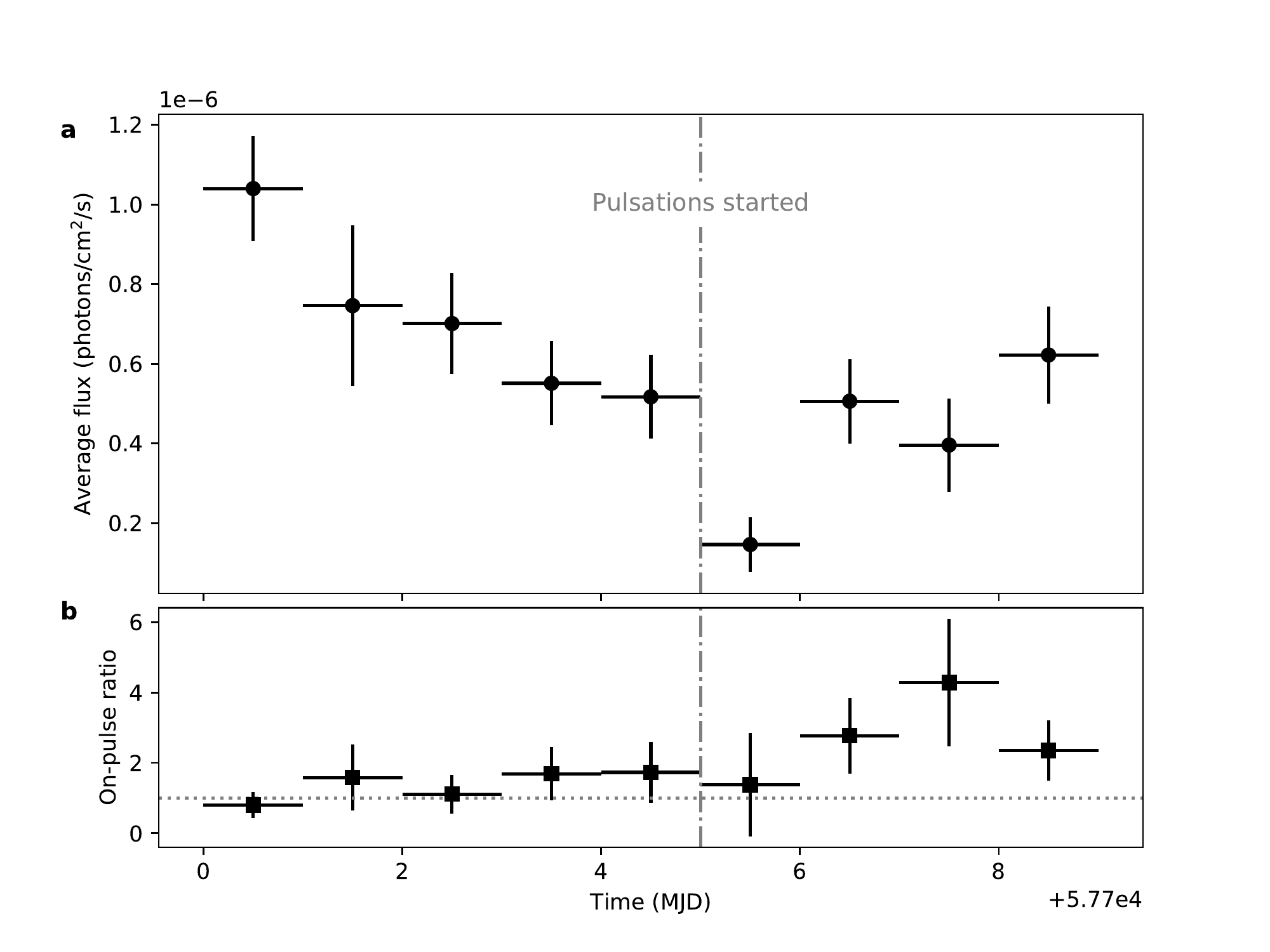}
\centering
\caption{The $\gamma$-ray light curve and the on-pulse flux ratio of \nova\ (i.e., the flux in the phase interval 0.45--0.55 over the flux in the entire phase interval).
Top panel (a): The phase-averaged daily $\gamma$-ray light curve of \nova\ (0.1--300~GeV) in MJD~57700.0--57709.0 obtained from \cite{2017NatAs...1..697L}. Bottom panel (b): The on-pulse-to-average $\gamma$-ray flux ratio over the same time period. The horizontal line indicates the flux ratio of 1, while the vertical line shows the starting date of the $\gamma$-ray pulsations. Both the average $\gamma$-ray light curve and the on-pulse contribution increased significantly since the pulsations started. All the reported errors are 1-$\sigma$ uncertainties.
}
\label{fig:ratio_lc}
\end{figure}

\section{Phase-Resolved \textit{Fermi}-LAT Likelihood Analysis}
Besides the aperture analysis, we performed a likelihood analysis with a much larger region of interest (ROI), $14\arcdeg\times14\arcdeg$, which is the largest square inside a circular region of radius $10\arcdeg$ centred on \nova, to measure the on/off-pulse $\gamma$-ray fluxes.
Except for the different ROI, the same data preparation and selection processes done by \texttt{gtselect} and \texttt{gtmktime} were performed as described in the previous section. In addition, the phase information computed by \texttt{gtbary} and \texttt{TEMPO2} was injected into the event file. 

The binned likelihood analysis method developed by the \textit{Fermi}-LAT Collaboration was adopted following the user manual provided in FSSC. For the likelihood fitting, a $\gamma$-ray emission model file for the field was constructed based on the fourth \textit{Fermi}-LAT source catalogue (4FGL; \citealt{2020ApJS..247...33A}). All the 4FGL $\gamma$-ray sources within 15\arcdeg\ from \nova\ were included in the model. In addition to the catalogued sources, two diffuse emission models, \texttt{gll\_iem\_v07} and \texttt{iso\_P8R3\_SOURCE\_V3\_v1} were added to account for the Galactic diffuse emission and the isotropic diffuse background, respectively.

The phase-resolved analysis focuses only on a 4-day interval from MJD~57705.0 to 57709.0. In this relatively short time, the quality of the observed $\gamma$-ray data is not sufficient for complex spectral modelling. A simple power-law model was therefore assumed. Since \nova\ was the dominant $\gamma$-ray source in the field \citep{2017NatAs...1..697L}, only the spectral parameters for \nova\ were allowed to vary and all other parameters in the model were fixed to their 4FGL values for simplicity. We defined the phase interval 0.45--0.55 as shown in Figure \ref{fig:phased_lc} to be the on-pulse phase and the time outside the phase range 0.4--0.6 as the off-pulse phase. There are short buffers between the on/off-pulse intervals to avoid any ambiguity during the transition. The likelihood model fittings were done by \texttt{gtlike} using the optimizer \texttt{NEWMINUIT}. The nova was significantly detected in both the on/off-pulse intervals with ${\rm TS}=92$ and ${\rm TS}=74$, respectively (i.e., the detection significance is approximately equal to $\sqrt{\rm TS}$ in the unit of $\sigma$). The on-pulse $\gamma$-ray emission can be characterised by a photon flux of $F_{\rm 0.1-300GeV}=(1.1\pm0.2)\times10^{-6}$\pflux\ with a photon index of $\Gamma=2.16\pm0.16$, while the parameters for the off-pulse emission are $F_{\rm 0.1-300GeV}=(3.3\pm0.7)\times10^{-7}$\pflux\ and $\Gamma=2.27\pm0.16$.

\subsection{On-Pulse-to-Average Flux Ratio Evolution}

The same approach was applied to extract the on-pulse $\gamma$-ray light curve for the understanding of the on-pulse flux contribution to the average emission over time. Daily bins from MJD~57700.0 to 57709.0 selected by \texttt{gtselect} were used. Given the short exposure time of each bin, the spectral model of \nova\ was further simplified by fixing the photon index to $\Gamma=2.16$, which is the best-fit value in the on-pulse spectral fitting. The nova was significantly detected in most of the dates (i.e., ${\rm TS} > 10$), and only marginally detected (${\rm TS}\approx4$) on MJD~57705, when the overall $\gamma$-ray flux dropped to the lowest flux level. 

The evolution of the on-pulse-to-average flux ratio (i.e., the flux in the phase interval 0.45--0.55 over the flux in the entire phase interval) is shown in Figure \ref{fig:ratio_lc}b, while Figure \ref{fig:ratio_lc}a shows the $\gamma$-ray light curve of \nova\ obtained in the previous study, in which a mysterious dip on MJD~57705 was reported by \cite{2017NatAs...1..697L}. Interestingly, the $\gamma$-ray pulsations started to appear on the same date, and we suspect that this is more than a coincidence. The on-pulse and the average fluxes were roughly the same before MJD~57705, but the on-pulse flux contribution increased by a factor of a few after then. This likely suggests an existence of an extra pulsed emission component that showed up around MJD~57705 to rebound the light curve and produce the emission dip.

\section{Discussion}

Time-domain observations have shown that white dwarf binaries can shine periodic pulses of light on time-scales from seconds to minutes, caused by the tilted magnetic fields of the compact stars. While the pulsation phenomenon has been observed from radio waves \citep{2016Natur.537..374M} to X-rays \citep{1980ApJ...240L.133P,2021ApJ...907..115T}, $\bm{\gamma}$-ray pulsations associated with a white dwarf have not hitherto been seen \citep{2016ApJ...832...35L,2020APh...12302488S}. The discovery of the 544.84-s $\gamma$-ray pulsations from \nova\ could be the first one seen in a white dwarf system.

\subsection{Is it an intermediate polar?}

While the unprecedented $\gamma$-ray pulsations is totally unexpected, it is reminiscent of the ``lighthouse effect'' for pulsars. In this context, the nova progenitor is likely a fast magnetic rotator, probably an intermediate polar \citep{1992A&A...257..548O}. Intermediate polars exploding as classical novae is not a novel idea, and there have been several well-known examples, such as DQ~Her \citep{1992A&A...257..548O}. For regular intermediate polars, the spin periods ($P_{\rm spin}$) are in the range of a few hundred to a few thousand seconds \citep{2004ApJ...614..349N}, which is in line with the 544.84-s signal of \nova. Unlike polars that have synchronized spin and orbital periods ($P_{\rm orb}$), the spin periods of intermediate polars are usually much shorter than the orbital periods \citep{2004ApJ...614..349N}, i.e., $0.25\gtrsim P_{\rm spin}/P_{\rm orb}\gtrsim 0.01$. The empirical relation suggests the orbital period of \nova\ to be 0.6--15 hours, which is consistent with the orbital period distribution of classical novae \citep{1997A&A...322..807D}.

\subsection{A bipolar wind model is unlikely}

Instead of light beams sweeping by Earth like a pulsar, we considered a case that strong bipolar nova winds \citep{2008ApJ...688..559R,2012A&A...545A..63C} from the magnetic poles of the white dwarf interact regularly with the matter deposited into the orbital plane by the early and slow ejection (slow ejecta hereafter; \citealt{2016MNRAS.455.4351P}) to create the $\gamma$-ray pulses. In the model, the axes of the white dwarf's spin, magnetic dipole, and orbital revolution all misalign. As the white dwarf rotates, $\gamma$-ray pulsations are formed as a natural consequence of the geometry.

However, in this model, the velocities of the polar winds ($v_p\sim1000$\kms) and slow ejecta ($v_s\sim100$\kms; \citealt{2017NatAs...1..697L}) have to be both very stable in the 4-day interval to create the observed pulsations. We can consider the observed $\sim$60-s on-pulse interval (Figure \ref{fig:phased_lc}) as the maximum phase shift between the first and the last observed pulses allowed by the data. Given that the nova was about 20 days old when it showed the $\gamma$-rays \citep{2017NatAs...1..697L}, the polar wind velocity change in the 4 days is limited by $\Delta v_p/v_p\approx ({\rm 60\,s/20\,d})\,v_p/v_s\approx0.035\%$. Similarly, by considering the fact that the $\gamma$-ray pulsations lasted for 4 days, $\Delta v_s/v_s\approx ({\rm 60\,s/4\,d})\,v_p/v_s\approx0.17\%$. Both conditions are rather extreme. Currently, $v_p/v_s\approx10$ is assumed, and a higher velocity ratio can slightly relieve the constraint. However, it is questionable whether such high velocity ratios are physical in such a classical nova system.

Besides, as the white dwarf in the binary is not stationary, the travel time for the polar wind from the white dwarf to the slow ejecta changes. If the orbit has a size of $1R_\sun$, the aforementioned change can be as large as $\sim$700 seconds, which will smear out the timing signal.

\section{Conclusion and Perspective}

We found a 544.84(7) periodic signal from \nova\ in $\gamma$-rays (significance $>4\sigma$), which is likely associated with the spin period of the white dwarf.
A bipolar wind model was discussed, but it is unlikely as the coherence of the signal will be lost unless the winds are extremely stable and the orbit is very compact.
Perhaps the coherent signal can be more easily understood, if the pulsed emission region is very close to the white dwarf.
However, it might contradict with the scenario of variable outflows that has been successful in explaining some observed properties in $\gamma$-ray novae \citep{2017NatAs...1..697L,2020NatAs...4..776A}.
Currently, we do not have a good explanation for the pulsation phenomenon and it remains open to interpretation.

From \nova\ and ASASSN-18fv, we have learnt that the optical and $\gamma$-ray light curves of a nova can closely track each other \citep{2017NatAs...1..697L,2020NatAs...4..776A}. Future detections of the optical pulsations associated with $\gamma$-rays are thus promising. Compared to $\gamma$-ray data, fast optical photometric observations would provide much improved ephemerides for nova pulsars, which could carry important information on the geometries of the binaries. With accurate ephemerides, pulsar gating analysis would also become possible to gain the detection sensitivity for $\gamma$-ray novae in GeV, or even TeV energies \citep{2016MNRAS.457.1786M}.

\begin{acknowledgements}
KLL is supported by the Ministry of Science and Technology of the Republic of China (Taiwan) through grant 110-2636-M-006-013, and he is a Yushan (Young) Scholar of the Ministry of Education of the Republic of China (Taiwan).
We acknowledge the use of public data from the \textit{Fermi}-LAT Data Server.
This work used high-performance computing facilities operated by the Center for Informatics and Computation in Astronomy (CICA) at National Tsing Hua University. This equipment was funded by the Ministry of Education of Taiwan, the Ministry of Science and Technology of Taiwan, and National Tsing Hua University.
\end{acknowledgements}

\facility{Fermi}

\bibliography{16ma_pulse}

\begin{thebibliography}{}
\expandafter\ifx\csname natexlab\endcsname\relax\def\natexlab#1{#1}\fi
\providecommand{\url}[1]{\href{#1}{#1}}
\providecommand{\dodoi}[1]{doi:~\href{http://doi.org/#1}{\nolinkurl{#1}}}
\providecommand{\doeprint}[1]{\href{http://ascl.net/#1}{\nolinkurl{http://ascl.net/#1}}}
\providecommand{\doarXiv}[1]{\href{https://arxiv.org/abs/#1}{\nolinkurl{https://arxiv.org/abs/#1}}}

\bibitem[{{Abdo} {et~al.}(2010){Abdo}, {Ackermann}, {Ajello}, {Atwood},
  {Baldini}, {Ballet}, {Barbiellini}, {Bastieri}, {Bechtol}, {Bellazzini}, \&
  et~al.}]{2010Sci...329..817A}
{Abdo}, A.~A., {Ackermann}, M., {Ajello}, M., {et~al.} 2010, Science, 329, 817,
  \dodoi{10.1126/science.1192537}

\bibitem[{{Abdollahi} {et~al.}(2020){Abdollahi}, {Acero}, {Ackermann},
  {Ajello}, {Atwood}, {Axelsson}, {Baldini}, {Ballet}, {Barbiellini},
  {Bastieri}, {Becerra Gonzalez}, {Bellazzini}, {Berretta}, {Bissaldi},
  {Blandford}, {Bloom}, {Bonino}, {Bottacini}, {Brandt}, {Bregeon}, {Bruel},
  {Buehler}, {Burnett}, {Buson}, {Cameron}, {Caputo}, {Caraveo}, {Casandjian},
  {Castro}, {Cavazzuti}, {Charles}, {Chaty}, {Chen}, {Cheung}, {Chiaro},
  {Ciprini}, {Cohen-Tanugi}, {Cominsky}, {Coronado-Bl{\'a}zquez}, {Costantin},
  {Cuoco}, {Cutini}, {D'Ammando}, {DeKlotz}, {de la Torre Luque}, {de Palma},
  {Desai}, {Digel}, {Di Lalla}, {Di Mauro}, {Di Venere}, {Dom{\'\i}nguez},
  {Dumora}, {Fana Dirirsa}, {Fegan}, {Ferrara}, {Franckowiak}, {Fukazawa},
  {Funk}, {Fusco}, {Gargano}, {Gasparrini}, {Giglietto}, {Giommi}, {Giordano},
  {Giroletti}, {Glanzman}, {Green}, {Grenier}, {Griffin}, {Grondin}, {Grove},
  {Guiriec}, {Harding}, {Hayashi}, {Hays}, {Hewitt}, {Horan},
  {J{\'o}hannesson}, {Johnson}, {Kamae}, {Kerr}, {Kocevski}, {Kovac'evic'},
  {Kuss}, {Landriu}, {Larsson}, {Latronico}, {Lemoine-Goumard}, {Li},
  {Liodakis}, {Longo}, {Loparco}, {Lott}, {Lovellette}, {Lubrano}, {Madejski},
  {Maldera}, {Malyshev}, {Manfreda}, {Marchesini}, {Marcotulli},
  {Mart{\'\i}-Devesa}, {Martin}, {Massaro}, {Mazziotta}, {McEnery}, {Mereu},
  {Meyer}, {Michelson}, {Mirabal}, {Mizuno}, {Monzani}, {Morselli},
  {Moskalenko}, {Negro}, {Nuss}, {Ojha}, {Omodei}, {Orienti}, {Orlando},
  {Ormes}, {Palatiello}, {Paliya}, {Paneque}, {Pei}, {Pe{\~n}a-Herazo},
  {Perkins}, {Persic}, {Pesce-Rollins}, {Petrosian}, {Petrov}, {Piron}, {Poon},
  {Porter}, {Principe}, {Rain{\`o}}, {Rando}, {Razzano}, {Razzaque}, {Reimer},
  {Reimer}, {Remy}, {Reposeur}, {Romani}, {Saz Parkinson}, {Schinzel},
  {Serini}, {Sgr{\`o}}, {Siskind}, {Smith}, {Spandre}, {Spinelli}, {Strong},
  {Suson}, {Tajima}, {Takahashi}, {Tak}, {Thayer}, {Thompson}, {Tibaldo},
  {Torres}, {Torresi}, {Valverde}, {Van Klaveren}, {van Zyl}, {Wood},
  {Yassine}, \& {Zaharijas}}]{2020ApJS..247...33A}
{Abdollahi}, S., {Acero}, F., {Ackermann}, M., {et~al.} 2020, \apjs, 247, 33,
  \dodoi{10.3847/1538-4365/ab6bcb}

\bibitem[{{Ackermann} {et~al.}(2014){Ackermann}, {Ajello}, {Albert}, {Baldini},
  {Ballet}, {Barbiellini}, {Bastieri}, {Bellazzini}, {Bissaldi}, {Blandford},
  {Bloom}, {Bottacini}, {Brandt}, {Bregeon}, {Bruel}, {Buehler}, {Buson},
  {Caliandro}, {Cameron}, {Caragiulo}, {Caraveo}, {Cavazzuti}, {Charles},
  {Chekhtman}, {Cheung}, {Chiang}, {Chiaro}, {Ciprini}, {Claus},
  {Cohen-Tanugi}, {Conrad}, {Corbel}, {D'Ammando}, {de Angelis}, {den Hartog},
  {de Palma}, {Dermer}, {Desiante}, {Digel}, {Di Venere}, {do Couto e Silva},
  {Donato}, {Drell}, {Drlica-Wagner}, {Favuzzi}, {Ferrara}, {Focke},
  {Franckowiak}, {Fuhrmann}, {Fukazawa}, {Fusco}, {Gargano}, {Gasparrini},
  {Germani}, {Giglietto}, {Giordano}, {Giroletti}, {Glanzman}, {Godfrey},
  {Grenier}, {Grove}, {Guiriec}, {Hadasch}, {Harding}, {Hayashida}, {Hays},
  {Hewitt}, {Hill}, {Hou}, {Jean}, {Jogler}, {J{\'o}hannesson}, {Johnson},
  {Johnson}, {Kerr}, {Kn{\"o}dlseder}, {Kuss}, {Larsson}, {Latronico},
  {Lemoine-Goumard}, {Longo}, {Loparco}, {Lott}, {Lovellette}, {Lubrano},
  {Manfreda}, {Martin}, {Massaro}, {Mayer}, {Mazziotta}, {McEnery},
  {Michelson}, {Mitthumsiri}, {Mizuno}, {Monzani}, {Morselli}, {Moskalenko},
  {Murgia}, {Nemmen}, {Nuss}, {Ohsugi}, {Omodei}, {Orienti}, {Orlando},
  {Ormes}, {Paneque}, {Panetta}, {Perkins}, {Pesce-Rollins}, {Piron}, {Pivato},
  {Porter}, {Rain{\`o}}, {Rando}, {Razzano}, {Razzaque}, {Reimer}, {Reimer},
  {Reposeur}, {Saz Parkinson}, {Schaal}, {Schulz}, {Sgr{\`o}}, {Siskind},
  {Spandre}, {Spinelli}, {Stawarz}, {Suson}, {Takahashi}, {Tanaka}, {Thayer},
  {Thayer}, {Thompson}, {Tibaldo}, {Tinivella}, {Torres}, {Tosti}, {Troja},
  {Uchiyama}, {Vianello}, {Winer}, {Wolff}, {Wood}, {Wood}, {Wood},
  {Charbonnel}, {Corbet}, {De Gennaro Aquino}, {Edlin}, {Mason}, {Schwarz},
  {Shore}, {Starrfield}, {Teyssier}, \& {Fermi-LAT
  Collaboration}}]{2014Sci...345..554A}
{Ackermann}, M., {Ajello}, M., {Albert}, A., {et~al.} 2014, Science, 345, 554,
  \dodoi{10.1126/science.1253947}

\bibitem[{{Aydi} {et~al.}(2020){Aydi}, {Sokolovsky}, {Chomiuk}, {Steinberg},
  {Li}, {Vurm}, {Metzger}, {Strader}, {Mukai}, {Pejcha}, {Shen}, {Wade},
  {Kuschnig}, {Moffat}, {Pablo}, {Pigulski}, {Popowicz}, {Weiss}, {Zwintz},
  {Izzo}, {Pollard}, {Handler}, {Ryder}, {Filipovi{\'c}}, {Alsaberi},
  {Manojlovi{\'c}}, {Lopes de Oliveira}, {Walter}, {Vallely}, {Buckley},
  {Brown}, {Harvey}, {Kawash}, {Kniazev}, {Kochanek}, {Linford},
  {Mikolajewska}, {Molaro}, {Orio}, {Page}, {Shappee}, \&
  {Sokoloski}}]{2020NatAs...4..776A}
{Aydi}, E., {Sokolovsky}, K.~V., {Chomiuk}, L., {et~al.} 2020, Nature
  Astronomy, 4, 776, \dodoi{10.1038/s41550-020-1070-y}

\bibitem[{{Bruel}(2019)}]{2019A&A...622A.108B}
{Bruel}, P. 2019, \aap, 622, A108, \dodoi{10.1051/0004-6361/201834555}

\bibitem[{{Chesneau} {et~al.}(2012){Chesneau}, {Lagadec}, {Otulakowska-Hypka},
  {Banerjee}, {Woodward}, {Harvey}, {Spang}, {Kervella}, {Millour}, {Nardetto},
  {Ashok}, {Barlow}, {Bode}, {Evans}, {Lynch}, {O'Brien}, {Rudy}, \&
  {Russell}}]{2012A&A...545A..63C}
{Chesneau}, O., {Lagadec}, E., {Otulakowska-Hypka}, M., {et~al.} 2012, \aap,
  545, A63, \dodoi{10.1051/0004-6361/201219825}

\bibitem[{{de Jager} {et~al.}(1989){de Jager}, {Raubenheimer}, \&
  {Swanepoel}}]{1989A&A...221..180D}
{de Jager}, O.~C., {Raubenheimer}, B.~C., \& {Swanepoel}, J.~W.~H. 1989, \aap,
  221, 180

\bibitem[{{Diaz} \& {Bruch}(1997)}]{1997A&A...322..807D}
{Diaz}, M.~P., \& {Bruch}, A. 1997, \aap, 322, 807

\bibitem[{{Drake} {et~al.}(2021){Drake}, {Ness}, {Page}, {Luna}, {Beardmore},
  {Orio}, {Osborne}, {Mr{\'o}z}, {Starrfield}, {Banerjee}, {Balman}, {Darnley},
  {Bhargava}, {Dewangan}, \& {Singh}}]{2021ApJ...922L..42D}
{Drake}, J.~J., {Ness}, J.-U., {Page}, K.~L., {et~al.} 2021, \apjl, 922, L42,
  \dodoi{10.3847/2041-8213/ac34fd}

\bibitem[{{Hobbs} {et~al.}(2006){Hobbs}, {Edwards}, \&
  {Manchester}}]{2006MNRAS.369..655H}
{Hobbs}, G.~B., {Edwards}, R.~T., \& {Manchester}, R.~N. 2006, \mnras, 369,
  655, \dodoi{10.1111/j.1365-2966.2006.10302.x}

\bibitem[{{Kerr}(2011)}]{2011ApJ...732...38K}
{Kerr}, M. 2011, \apj, 732, 38, \dodoi{10.1088/0004-637X/732/1/38}

\bibitem[{{Li} {et~al.}(2016){Li}, {Torres}, {Rea}, {de O{\~n}a Wilhelmi},
  {Papitto}, {Hou}, \& {Mauche}}]{2016ApJ...832...35L}
{Li}, J., {Torres}, D.~F., {Rea}, N., {et~al.} 2016, \apj, 832, 35,
  \dodoi{10.3847/0004-637X/832/1/35}

\bibitem[{{Li} {et~al.}(2017){Li}, {Metzger}, {Chomiuk}, {Vurm}, {Strader},
  {Finzell}, {Beloborodov}, {Nelson}, {Shappee}, {Kochanek}, {Prieto}, {Kafka},
  {Holoien}, {Thompson}, {Luckas}, \& {Itoh}}]{2017NatAs...1..697L}
{Li}, K.-L., {Metzger}, B.~D., {Chomiuk}, L., {et~al.} 2017, Nature Astronomy,
  1, 697, \dodoi{10.1038/s41550-017-0222-1}

\bibitem[{{Maccarone} {et~al.}(2021){Maccarone}, {Beardmore}, {Mukai}, {Page},
  {Pichardo Marcano}, \& {Rivera Sandoval}}]{2021ATel14776....1M}
{Maccarone}, T.~J., {Beardmore}, A., {Mukai}, K., {et~al.} 2021, The
  Astronomer's Telegram, 14776, 1

\bibitem[{{Marsh} {et~al.}(2016){Marsh}, {G{\"a}nsicke}, {H{\"u}mmerich},
  {Hambsch}, {Bernhard}, {Lloyd}, {Breedt}, {Stanway}, {Steeghs}, {Parsons},
  {Toloza}, {Schreiber}, {Jonker}, {van Roestel}, {Kupfer}, {Pala}, {Dhillon},
  {Hardy}, {Littlefair}, {Aungwerojwit}, {Arjyotha}, {Koester}, {Bochinski},
  {Haswell}, {Frank}, \& {Wheatley}}]{2016Natur.537..374M}
{Marsh}, T.~R., {G{\"a}nsicke}, B.~T., {H{\"u}mmerich}, S., {et~al.} 2016,
  \nat, 537, 374, \dodoi{10.1038/nature18620}

\bibitem[{{Metzger} {et~al.}(2016){Metzger}, {Caprioli}, {Vurm}, {Beloborodov},
  {Bartos}, \& {Vlasov}}]{2016MNRAS.457.1786M}
{Metzger}, B.~D., {Caprioli}, D., {Vurm}, I., {et~al.} 2016, \mnras, 457, 1786,
  \dodoi{10.1093/mnras/stw123}

\bibitem[{{Mroz} {et~al.}(2021){Mroz}, {Burdge}, {Roestel}, {Prince}, {Kong},
  \& {Li}}]{2021ATel14720....1M}
{Mroz}, P., {Burdge}, K., {Roestel}, J.~v., {et~al.} 2021, The Astronomer's
  Telegram, 14720, 1

\bibitem[{{Norton} {et~al.}(2004){Norton}, {Wynn}, \&
  {Somerscales}}]{2004ApJ...614..349N}
{Norton}, A.~J., {Wynn}, G.~A., \& {Somerscales}, R.~V. 2004, \apj, 614, 349,
  \dodoi{10.1086/423333}

\bibitem[{{Orio} {et~al.}(1992){Orio}, {Trussoni}, \&
  {Oegelman}}]{1992A&A...257..548O}
{Orio}, M., {Trussoni}, E., \& {Oegelman}, H. 1992, \aap, 257, 548

\bibitem[{{Patterson} {et~al.}(1980){Patterson}, {Branch}, {Chincarini}, \&
  {Robinson}}]{1980ApJ...240L.133P}
{Patterson}, J., {Branch}, D., {Chincarini}, G., \& {Robinson}, E.~L. 1980,
  \apjl, 240, L133, \dodoi{10.1086/183339}

\bibitem[{{Pejcha} {et~al.}(2016){Pejcha}, {Metzger}, \&
  {Tomida}}]{2016MNRAS.455.4351P}
{Pejcha}, O., {Metzger}, B.~D., \& {Tomida}, K. 2016, \mnras, 455, 4351,
  \dodoi{10.1093/mnras/stv2592}

\bibitem[{{Rupen} {et~al.}(2008){Rupen}, {Mioduszewski}, \&
  {Sokoloski}}]{2008ApJ...688..559R}
{Rupen}, M.~P., {Mioduszewski}, A.~J., \& {Sokoloski}, J.~L. 2008, \apj, 688,
  559, \dodoi{10.1086/525555}

\bibitem[{{Shappee} {et~al.}(2014){Shappee}, {Prieto}, {Grupe}, {Kochanek},
  {Stanek}, {De Rosa}, {Mathur}, {Zu}, {Peterson}, {Pogge}, {Komossa}, {Im},
  {Jencson}, {Holoien}, {Basu}, {Beacom}, {Szczygie{\l}}, {Brimacombe},
  {Adams}, {Campillay}, {Choi}, {Contreras}, {Dietrich}, {Dubberley},
  {Elphick}, {Foale}, {Giustini}, {Gonzalez}, {Hawkins}, {Howell}, {Hsiao},
  {Koss}, {Leighly}, {Morrell}, {Mudd}, {Mullins}, {Nugent}, {Parrent},
  {Phillips}, {Pojmanski}, {Rosing}, {Ross}, {Sand}, {Terndrup}, {Valenti},
  {Walker}, \& {Yoon}}]{2014ApJ...788...48S}
{Shappee}, B.~J., {Prieto}, J.~L., {Grupe}, D., {et~al.} 2014, \apj, 788, 48,
  \dodoi{10.1088/0004-637X/788/1/48}

\bibitem[{{Singh} {et~al.}(2020){Singh}, {Meintjes}, {Kaplan}, {Ramamonjisoa},
  \& {Sahayanathan}}]{2020APh...12302488S}
{Singh}, K.~K., {Meintjes}, P.~J., {Kaplan}, Q., {Ramamonjisoa}, F.~A., \&
  {Sahayanathan}, S. 2020, Astroparticle Physics, 123, 102488,
  \dodoi{10.1016/j.astropartphys.2020.102488}

\bibitem[{{Stanek} {et~al.}(2016){Stanek}, {Kochanek}, {Brown}, {Holoien},
  {Shields}, {Shappee}, {Prieto}, {Bersier}, {Dong}, {Bose}, {Chen}, {Chomiuk},
  {Strader}, \& {Brimacombe}}]{2016ATel.9669....1S}
{Stanek}, K.~Z., {Kochanek}, C.~S., {Brown}, J.~S., {et~al.} 2016, The
  Astronomer's Telegram, 9669, 1

\bibitem[{{Takata} {et~al.}(2021){Takata}, {Wang}, {Wang}, {Lin}, {Hu}, {Li},
  \& {Kong}}]{2021ApJ...907..115T}
{Takata}, J., {Wang}, X.~F., {Wang}, H.~H., {et~al.} 2021, \apj, 907, 115,
  \dodoi{10.3847/1538-4357/abd0f8}

\end{thebibliography}

\end{document}